\documentclass[pra,twocolumn,nofootinbib,showpacs,showkeys,preprintnumbers,amsmath,amssymb]{revtex4-2}

\usepackage{graphicx}
\usepackage{dcolumn}
\usepackage{bm}
\usepackage[usenames]{color}
\usepackage{color}
\usepackage[colorlinks={true}]{hyperref}
\hypersetup{colorlinks=true,linkcolor=red,citecolor=blue,urlcolor=blue}
\usepackage{graphicx,epsfig}
\usepackage{orcidlink} 
\usepackage{pstricks}

\begin{document}
\title{
Quantum correlations versus spin magnitude: Transition to the classical limit
}

\author{M.~A.~Yurischev\,\orcidlink{0000-0003-1719-3884}}
\email{yur@itp.ac.ru}
\affiliation{
Federal Research Center of Problems of Chemical Physics and Medicinal Chemistry,
Russian Academy of Sciences, Chernogolovka 142432, Moscow Region, Russia
}
\author{E.~I.~Kuznetsova\,\orcidlink{0000-0002-0053-0023}}
\email{kuznets@icp.ac.ru}
\affiliation{
Federal Research Center of Problems of Chemical Physics and Medicinal Chemistry,
Russian Academy of Sciences, Chernogolovka 142432, Moscow Region, Russia
}

\author{Saeed~Haddadi\,\orcidlink{0000-0002-1596-0763}}
\email{haddadi@ipm.ir}
\affiliation{School of Particles and Accelerators, Institute for Research in\\
Fundamental Sciences (IPM), P.O. Box 19395-5531, Tehran, Iran}

\date{\today}

\begin{abstract}
Quantum-classical transitions have long attracted much attention.
We study such transitions in quantum spin-($j$,1/2) systems at thermal equilibrium.
Unlike the previous paper [\href{https://doi.org/10.1103/PhysRevA.73.064302}{Phys. Rev. A {\bf 73}, 064302 (2006)}],
it is found that the threshold temperature of quantum
entanglement decreases with increasing spin $j$ and completely disappears in the limit
$j\to\infty$.
In the ground state of systems with highly symmetric interactions, the discord-type
quantum correlations can exist even for arbitrarily large spin.
Such correlations turn out to be unstable and are destroyed by small perturbations that
violate the symmetry of the Hamiltonian.
The stable quantum correlations gradually degrade as the spin $j$ grows and eventually
vanish when the classical limit is reached.
\end{abstract}

\medskip 
\pacs{03.65.-w, 03.65.Ud, 03.67.-a, 75.10.Jm} 

\keywords{Qudit-qubit system, Axial symmetry group U(1), Local quantum uncertainty,
Local quantum Fisher information} %

\maketitle

\section{Introduction}
\label{sect:Intro}
According to the correspondence principle, proposed by Bohr in 1920, a quantum system
at large quantum numbers should transform into a classical one \cite{B20,L84}.
However, classical theory does not automatically follow from quantum theory.
These points can be illustrated with a heuristic example.

Consider the transition from Brillouin's quantum theory of paramagnetism to Langevin's
classical theory.
Let there be a single spin $j$ in an external magnetic field $B$ applied along $z$-axis.
The Zeeman Hamiltonian reads
\begin{equation}
   \label{eq:H_Z}
   {\cal H}=-\mu BS_z,
\end{equation}
where $S_z={\rm diag}[-j,-j+1,\ldots,j]$ and $\mu=g\mu_B$, in which $g$ is the
$g$-factor and $\mu_B$ the Bohr magneton (electronic or nucleic).
The average magnetization of $N$ spins at the temperature $T$ (in energy units) is
given as
\begin{equation}
   \label{eq:Brill}
   M=N\mu{\rm Tr}\big(S_ze^{-{\cal H}/T}\big)/{\rm Tr}e^{-{\cal H}/T}=N\mu j{\cal B}_j(\mu jB/T),
\end{equation}
where
${\cal B}_j(x)=\frac{2j+1}{2j}\coth{\big(\frac{2j+1}{2j}x\big)}-\frac{1}{2j}\coth{\big(\frac{x}{2j}\big)}$
is called the Brillouin function \cite{VV32,V71,K05,B99}.
For large quantum number $j\to\infty$, the Brillouin function is reduced to the
Langevin one, $B_\infty(x)=L(x)=\coth x-1/x$.
Equation~(\ref{eq:Brill}) will reproduce the Langevin magnetization
$M=N\mu_0L(\mu_0B/T)$, if the {\em additional} condition is performed, namely
$j\mu=\mu_0$, where $\mu_0$ is the classical magnetic moment, the magnitude of which is
fixed, but it can have arbitrary orientations in the three-dimensional space
(Langevin's classical spin).
In other words, this condition means that the spin operator $S_z$ in Eq.~(\ref{eq:H_Z})
is now normalized, ${\cal H}=-\mu_0 BS_z/j$, and the diagonal matrix $S_z/j$ becomes a
continuous variable on the segment $[-1,1]$ as the spin increases.\\

The idea of rescaling spin operators was then used to prove that the quantum Heisenberg
model transforms into the classical Heisenberg system in the limit of large spins
\cite{F64,St71,ML71,M75,CS90,MSL99,N04}.
This opened the way to the study of magnetic phase transitions in Heisenberg models at
finite temperatures.

It is worth noting that spin normalization is widely used in high-spin Ising models
\cite{JGE96,BC02,BCG03,Y06,Y06a,AB21}.
However, normalization here serves a different purpose.
The division by $j$ is done to conserve the energy scale across the different
spin-$j$ models and thereby make meaningful temperature comparisons between them
\cite{AB21}.

On the other hand, normalization is not applied in solid state physics \cite{K05},
magnetochemistry \cite{C86} and so on, since the magnetic moments of the ions or atoms
used are small and the problem of studying systems with huge spins does not arise here.
There are also various single-molecule magnets \cite{IK00,BW08,BG15,MW21,CKM21}.
The total spin in the molecules of such materials can reach values of 10 and more, but
the goals of molecular magnetism are also different \cite{BG15}.

It is also worth mentioning Rydberg atoms.
The principal quantum number in them can reach values of $n\sim1000$ and more
\cite{FHSDF95}.
The possibilities of obtaining states of such macroscopic objects with a large orbital
angular momentum ($l\sim n$) are considered \cite{R84,RMM16,BHGJF20}.
Quantum information processing based on Rydberg atoms is currently being intensively
discussed \cite{WLT21}.

The transformation of quantum properties of matter into the visible classical
properties of the world around us has been of genuine interest since the emergence of
quantum theory \cite{S07,CFIM12,K22,LN23,ALVGP24,SRS24}.
Various possibilities for such transitions are discussed: gradual modification of the
quantum system with an increase in its size, the emergence of classicality through an
abrupt transition, or the system continues to remain quantum even when reaching
macroscopic scales (for example, the phenomena of superfluidity and superconductivity).
An important role in establishing the mechanisms of quantum-to-classical transitions
belongs to specific models that allow a rigorous description or numerical simulation.

Hamiltonian~(\ref{eq:H_Z}) is obviously classical, it corresponds to the
high-dimensional Ising spin.
It is therefore interesting to find a non-trivial example of a coupled-spin quantum
system that allows a detailed description of the transition to the classical limit.
Here we have settled on a two-particle system composed of a spin of arbitrary value
$j$ and a spin of 1/2.
The qudit-qubit system is attractive because it lends itself to analytical
calculations.

The highly symmetric SU(2)-case of the qubit-qudit model was discussed in many works.
First, the structure of SU(2)-invariant states was considered in Refs.~\cite{S03,S05}.
These states are invariant under uniform rotation of both spins, which means that the
density matrix commutes with all components of the total spin.
For this case, various important nonclassical correlations, such as quantum
entanglement in the form of negativity \cite{S03,S05,WW06}, quantum entanglement of
formation (EoF) \cite{MC08}, relative entropy of entanglement \cite{WW08}, quantum
discord (QD) \cite{CG13}, one-way deficit \cite{WMFW14} and local quantum uncertainty
(LQU) \cite{FA15} have been found.
Moreover, the Holevo quantity has been also obtained in \cite{WGFW22}.
Note in passing that this list does not include local quantum Fisher information
(LQFI); below we will fill this gap by deriving Eq.~(\ref{eq:Fxxx-vsF}).

The SU(2)-invariant state is one-parameter and corresponds to the thermal state of only
the pure isotropic Heisenberg model.
Quantum correlations in qubit-qudit systems with more general interactions were also
discussed in a number of papers.
For example, quantum entanglement, quantified by negativity, was investigated in
Refs.~\cite{LRKL12,VS21,VST22}, QD in some extended X states was
presented in \cite{VR12}, high-temperature dynamics of quantum and classical
correlations was examined in Refs.~\cite{Z13r,Z13}, and ground-state entropic
entanglement was analyzed in \cite{ML13}.
In the paper \cite{YH23}, exact formulas of LQU and LQFI for general qubit-qubit
X states were provided.
Moreover, some studies have been carried out on quantum correlations in hybrid
qubit-qutrit systems using negativity, as well as LQU and LQFI \cite{BARDAH23,YHG25}.
Recently, analytical expressions for LQU and LQFI were found in the case of
axisymmetric states for $2\otimes d$ systems with arbitrary dimension $d =2j+1$ of the
second subsystem \cite{HKY25}.

Closed forms of LQU and LQFI for most general $2\otimes d$ systems have been evaluated
in Refs.~\cite{GTA13,GSGTFSSOA14}.
Unfortunately, the derived formulas include the need to solve algebraic equations of
the third degree.
This circumstance poses a serious obstacle to the practical calculation and study of
quantum correlations.
Therefore, it is much better to reduce the number of different interactions in the
system in order to obtain significantly simpler formulas for quantum correlations.

In this paper, we impose axial symmetry on the system, i.e. we assume that the
Hamiltonian (or the density matrix) commutes with only one component ($z$-th, without
loss of generality) of the total spin.
This constraint preserves a large number of physically important interactions
(Heisenberg XXZ coupling, inhomogeneous external magnetic fields, etc.).
On the other hand, both LQU and LQFI are now expressed only in terms of {\em square}
roots.
In Appendix~\ref{sec:append}, we rederive the important formulas for all four branches
of the correlations LQU and LQFI using a new ordering for subsystem spaces, namely
$C^d\times C^2$ instead of $C^2\times C^d$.
This permutation of subspaces leads to an explicit block-diagonal structure of both the
Hamiltonian and the density matrix.

Using the obtained formulas for discord-like correlations, we conduct extensive studies
of their properties both at finite and at zero temperatures.
The main focus is on revealing the behavior of quantum correlations at large spins.

The remainder of the article is organized as follows.
The next section provides a detailed introduction to the quantum correlation measures
LQU and LQFI.
We describe the axially invariant model in Sec.~\ref{sec:ASH}.
The main results are presented in Sec.~\ref{sec:Res}.
The study is concluded in Sec.~\ref{sec:Concl}.

\section{
Preliminaries
}
\label{sect:prelim}
Here, we recall some definitions and equations related to LQU and LQFI.

\subsection{
Local quantum uncertainty
}
\label{sect:LQU}
The total uncertainty of an observable $H$ in the quantum-mechanical state $\rho$ is
usually expressed by the variance
\begin{equation}
   \label{eq:var}
   {\rm Var}(\rho,H)=\langle H^2\rangle_{\rho}-\langle H\rangle_{\rho}^2.
\end{equation}
On the other hand, the quantum contribution to the total statistical error may be
reliable quantify via the Wigner-Yanase skew information \cite{WY63,L03,L03a}
\begin{equation}
   \label{eq:WY}
   {\cal I}_{\text{WY}}(\rho,H)=-\frac{1}{2}{\rm Tr}[\sqrt{\rho},H]^2,
\end{equation}
where $[.,.]$ denotes the commutator.
Notice that the skew information (\ref{eq:WY}) is not grater than the variance
(\ref{eq:var}), namely
\begin{equation}
   \label{eq:WY-V}
   {\cal I}_{\text{WY}}(\rho,H)\le{\rm Var}(\rho,H),
\end{equation}
where the equality is achieved for pure states when classical ignorance does not occur
\cite{GTA13}.
This makes it possible to introduce a discord-type measure $\cal U$ (also called LQU)
of quantum correlations in any bipartite system $AB$ as follows
\begin{equation}
   \label{eq:Udef}
   {\cal U}(\rho)=\min_{H_B}{\cal I}_{\text{WY}}(\rho,H_B),
\end{equation}
in which the minimum is taken over all local observables $H_B$ on the subsystem $B$.
It is worth mentioning that the LQU is a genuine quantifier of quantum correlations,
and it has been shown that the LQU meets all the physical conditions of a measure of
quantum correlations.

\subsection{
Local quantum Fisher information
}
\label{sect:LQFI}
Quantum Fisher information (QFI) is the cornerstone of quantum estimation theory
\cite{H76,H82,BC94,LYLW20}.
Suppose there is unitary evolution of the quantum state
$\varrho=e^{iH\epsilon}\rho e^{-iH\epsilon}$ with some observable $H$, and
the parameter $\epsilon$ wanted to be estimated.
The QFI can be defined as follows
\begin{equation}
   \label{eq:QFI}
   {\cal I}_F(\varrho, H)=\frac{1}{4}{\rm Tr}(\varrho L_\epsilon^2),
\end{equation}
where $L_\epsilon$ is the symmetric logarithmic derivative (SLD) operator satisfying
the Lyapunov equation (known from control theory)
\begin{equation}
   \label{eq:SLD}
   \frac{\partial\varrho}{\partial\epsilon}=\frac{1}{2}(\varrho L_\epsilon+L_\epsilon\varrho).
\end{equation}
Then, the accuracy of the estimation of $\epsilon$ is limited by the quantum
Rao-Cram$\rm{\acute e}$r inequality \cite{H82}
\begin{equation}
   \label{eq:RC}
   \Delta\epsilon\ge\frac{1}{\sqrt{n {\cal I}_F(\varrho,H)}},
\end{equation}
where $n$ is the number of measurements.

Another application of QFI is to use it as a quantifier of quantum correlations
\cite{GSGTFSSOA14}.
It is showed that the quantity
\begin{equation}
   \label{eq:Fdef}
   {\cal F}(\rho)=\min_{H_B}{\cal I}_F(\rho,H_B),
\end{equation}
where $H_B$ (as in the case of LQU) is a local observable, again satisfies all
the necessary criteria to be defined as a measure of quantum correlations.
Note that here the (local) QFI is minimized, unlike the case of the lower bound
(\ref{eq:RC}).
The measure $\cal F$ is called interferometric power \cite{GSGTFSSOA14} or LQFI
\cite{B14,KLKW18}.

The authors of Refs.~\cite{GTA13,GSGTFSSOA14} were able to perform the required
minimizations in Eqs.~(\ref{eq:Udef}) and (\ref{eq:Fdef}) only for the system
consisting of a qudit and a qubit, and obtained both measures LQU and LQFI in closed
form.

\section{Axially symmetric Hamiltonian and Gibbs density matrix}
\label{sec:ASH}
Let us consider two-particle system $(j,1/2)$ composed of a spin $j$ $(j=1/2,1,\dots)$
and a spin 1/2.
We restrict ourselves to the Hamiltonians which are invariant under transformations of
the axial symmetry group U(1).
The group U(1) consists of rotations $R_z(\phi)=\exp(-i\phi{\cal S}_z)$ around the
$z$-axis on angles $\phi\in[0,\pi)$.
This means that the Hamiltonian commutes with the $z$-component of the total spin,
$[{\cal H},{\cal S}_z]=0$.
The component ${\cal S}_z$ of total spin is given by
\begin{eqnarray}
   \label{eq:Sz_total}
   &&{\cal S}_z=S_z\otimes I_2 +I_{2j+1}\otimes s_z={\rm diag}\,[j+1/2,j-1/2,j
   \nonumber\\
   &&-1/2,\ldots,-j+1/2,-j+1/2,-j-1/2].
\end{eqnarray}
Here, $S_z={\rm diag}[j,j-1,\ldots,-j]$, $s_z=\sigma_z/2$ ($\sigma_z$ is the Pauli $z$
matrix) and $I_2$ and $I_{2j+1}$ are the identity operators of the second and
($2j+1$)-th orders, respectively.

The most general Hermitian matrix which commutes with the $z$-component of total spin
(\ref{eq:Sz_total}) can be written as
\begin{equation}
   \label{eq:Has}
   {\cal H}=
	 \left(
      \begin{array}{cccccccccccc}
      E_0&\ &\ &\ &\ &\ &\ &\ &\ &\ &\ &\ \\
      \ &h_1&g_1&\ &\ &\ &\ &\ &\ &\ &\ &\ \\
      \ &g_1^*&h^\prime_1&\ &\ &\ &\ &\ &\ &\ \ &\ &\ \\
      \ &\ &\ &\ddots&\ &\ &\ &\ &\ &\ &\ &\ \\
      \ &\ &\ &\ &h_k&g_k&\ &\ &\ &\ &\ &\ \\
      \ &\ &\ &\ &g_k^*&h^\prime_k&\ &\ &\ &\ &\ &\ \\
      \ &\ &\ &\ &\ &\ &\ &\ddots&\ &\ &\ &\ \\
      \ &\ &\ &\ &\ &\ &\ &\ &\ &h_{2j}&g_{2j}&\ \\
      \ &\ &\ &\ &\ &\ &\ &\ &\ &g_{2j}^*&h^\prime_{2j}&\ \\
			\ &\ &\ &\ &\ &\ &\ &\ &\ &\ &\ &E_{4j+1}
      \end{array}
   \right).
\end{equation}
This matrix consists of two $1\times1$ subblocks and $2j$ subblocks of second orders.

The energy spectrum of the Hamiltonian (\ref{eq:Has}) consists of the levels $E_ 0$,
$E_{4j+1}$ and $2j$ additional pairs
\begin{equation}
   \label{eq:Ei_1}
   E_k^{(1,2)}=\frac{1}{2}\big(h_k+h^\prime_k\pm R_k\big),
\end{equation}
where
\begin{equation}
   \label{eq:Rk_2}
   R_k=\sqrt{(h_k-h^\prime_k)^2+4|g_k|^2}
\end{equation}
and $k=1,\ldots,2j$.

We are interested in a system that is in thermal equilibrium with the environment.
Its partition function
$Z=\sum_n\exp(-E_n/T)$ is given as
\begin{equation}
   \label{eq:Z}
   Z=e^{-E_0/T}+e^{-E_{4j+1}/T}+2\sum_{k=1}^{2j}\cosh{\Big(\frac{R_k}{2T}\Big)}e^{-(h_k+h^\prime_k)/2T}.
\end{equation}
The Gibbs density matrix is equal to
\begin{equation}
   \label{eq:rhoG0_3}
   \rho=\frac{1}{Z}\exp(-{\cal H}/T).
\end{equation}
Statistical weights, i.e. eigenvalues of this density matrix, would be presented as
\begin{eqnarray}
   \label{eq:p_i_3}
   &&p_0=\frac{1}{Z}e^{-E_0/T},\quad
	 p_k=\frac{1}{Z}e^{-E^{(1)}_k/T},
   \nonumber\\
   &&q_k=\frac{1}{Z}e^{-E^{(2)}_k/T},\quad p_{4j+1}=\frac{1}{Z}e^{-E_{4j+1}/T}.
\end{eqnarray}

On the other hand, since $\cal H$ has the axially symmetrical form, the functional
relationship (\ref{eq:rhoG0_3}) also leads to the density matrix of the same structure,
see Eq.~(\ref{eq:rho}) in Appendix~\ref{sec:append}.
If the dimension of the matrix $\cal H$ is finite, $\rho$ can be calculated manually or
obtained on a digital machine by direct symbolic (analytical) calculations.
After this, the results can be extended by induction to systems with arbitrary spin
$j$.
This allows us to get all non-zero matrix elements of the two-by-two blocks of the
density matrix (\ref{eq:rho}) through the parameters of Hamiltonian:
\begin{eqnarray}
   \label{eq:akbkuk_3}
   &&a_k=\frac{1}{Z}\Big(\cosh{\frac{R_k}{2T}}+\frac{h^\prime_k-h_k}{R_k}\sinh{\frac{R_k}{2T}}\Big)e^{-(h_k+h^\prime_k)/2T},
   \nonumber\\
   &&b_k=\frac{1}{Z}\Big(\cosh{\frac{R_k}{2T}}-\frac{h^\prime_k-h_k}{R_k}\sinh{\frac{R_k}{2T}}\Big)e^{-(h_k+h^\prime_k)/2T},
   \nonumber\\
	 &&u_k=-\frac{2g_k}{ZR_k}\sinh{\frac{R_k}{2T}}e^{-(h_k+h^\prime_k)/2T}
\end{eqnarray}
with $k=1,\ldots,2j$.
In turn, these expressions open up an alternative opportunity to find the
eigenvalues of the Gibbs density matrix using Eq.~(\ref{eq:pkqk}).

Now, we move on to discussing the properties of quantum correlations of specific
axially symmetric systems.

\section{Results and discussion}
\label{sec:Res}


\subsection{Heisenberg XXX model}
\label{subsec:XXX}
We start the study of bipartite system $AB$ with the pure isotropic Heisenberg XXX
Hamiltonian
\begin{equation}
   \label{eq:Hxxx}
   {\cal H}=J_0\frac{{\bf S}}{|{\bf S}|}\cdot{\bf s}
   =\frac{J_0}{2\sqrt{j(j+1)}}(S_x\sigma_x+S_y\sigma_y+S_z\sigma_z),
\end{equation}
where $\bf S$ and $\bf s$ are the vectors of spin operators of magnitudes $j$ and 1/2,
respectively; $\sigma_x$, $\sigma_y$ and $\sigma_z$ are the Pauli matrices.
Thus, the spin operators $S_i$ ($i=x,y,z$) are normalized to the spin length
$|{\bf S}|=\sqrt{j(j+1)}\approx j$, but $s_i=\sigma_i/2$
are not, since the spin of particle $B$ is fixed and does not tend to infinity.
The spin operators satisfy to the commutation relation $[S_i,S_j]=i\epsilon_{ijk}S_k$,
where $\epsilon_{ijk}$ is the Levi-Civita completely antisymmetric tensor with
$\epsilon_{xyz}=1$.
In the classical limit $j\to\infty$, the normalized spin operators $S_x/|{\bf S}|$,
$S_y/|{\bf S}|$ and $S_z/|{\bf S}|$ commute with each other.

Using the standard form of the spin-$j$ and spin-1/2 operators (see, for example,
\cite{B99}), we obtain from (\ref{eq:Hxxx}) that
\begin{widetext}
\begin{equation}
   \label{eq:Hopen}
   {\cal H}=\frac{J_0}{2\sqrt{j(j+1)}}
	 \left(
      \begin{array}{cccccccccccc}
      j&\ &\ &\ &\ &\ &\ &\ &\ &\ &\ &\ \\
      \ &-j&\sqrt{2j}&\ &\ &\ &\ &\ &\ &\ &\ &\ \\
      \ &\sqrt{2j}&j-1&\ &\ &\ &\ &\ &\ &\ \ &\ &\ \\
      \ &\ &\ &\ddots&\ &\ &\ &\ &\ &\ &\ &\ \\
      \ &\ &\ &\ &-j+k-1&\sqrt{k(2j-k+1)}&\ &\ &\ &\ &\ &\ \\
      \ &\ &\ &\ &\sqrt{k(2j-k+1)}&j-k&\ &\ &\ &\ &\ &\ \\
      \ &\ &\ &\ &\ &\ &\ &\ddots&\ &\ &\ &\ \\
      \ &\ &\ &\ &\ &\ &\ &\ &\ &j-1&\sqrt{2j}&\ \\
      \ &\ &\ &\ &\ &\ &\ &\ &\ &\sqrt{2j}&-j&\ \\
			\ &\ &\ &\ &\ &\ &\ &\ &\ &\ &\ &j
      \end{array}
   \right).
\end{equation}
\end{widetext}
This matrix has an axially symmetrical structure (\ref{eq:Has}).

The energy spectrum of the system (\ref{eq:Hopen}) consist of two different levels
\begin{equation}
   \label{eq:E12}
   E_1=\frac{jJ_0}{2\sqrt{j(j+1)}},\qquad
	 E_2=-\frac{(j+1)J_0}{2\sqrt{j(j+1)}}
\end{equation}
with degeneracies $2(j+1)$ and $2j$, respectively.
The partition function (\ref{eq:Z}) is reduced to
\begin{equation}
   \label{eq:Zxxx}
   Z=2(j+1)\exp\bigg[-\frac{jJ_0}{2\sqrt{j(j+1)}T}\bigg]+2j\exp\bigg[\frac{(j+1)J_0}{2\sqrt{j(j+1)}T}\bigg].
\end{equation}
The Gibbs density matrix (\ref{eq:rhoG0_3}) is written as
\begin{widetext}
\begin{equation}
   \label{eq:rhoGxxx}
   \rho=\frac{1}{2(2j+1)(1+j(1+w))}
	 \left(
      \begin{array}{cccccccc}
      2j+1&\ &\ &\ &\ &\ &\ &\ \\
      \ &\ddots&\ &\ &\ &\ &\ &\ \\
      \ &\ &k+(2j+1-k)w&(w-1)\sqrt{k(2j+1-k)}&\ &\ &\ &\ \\
      \ &\ &(w-1)\sqrt{k(2j+1-k)}&2j+1+k(w-1)&\ &\ &\ &\ \\
      \ &\ &\ &\ &\ddots&\ &\ &\ \\
			\ &\ &\ &\ &\ &\ &\ &2j+1
      \end{array}
   \right),
\end{equation}
\end{widetext}
where
\begin{equation}
   \label{eq:w}
   w=\exp{\Bigg(\frac{(2j+1)J_0}{2T\sqrt{j(j+1)}}\Bigg)}.
\end{equation}
The state (\ref{eq:rhoGxxx}) belongs to the class of axially symmetric states defined
by Eq.~(\ref{eq:rho}).
Comparison Eqs.~(\ref{eq:rhoGxxx}) and (\ref{eq:rho}) allows us to establish that the
matrix elements of the density matrix are given as
\begin{equation}
   \label{eq:xxx_a_k}
   a_k=\frac{k+(2j+1-k)w}{2(2j+1)[1+j(1+w)]},
\end{equation}
\begin{equation}
   \label{eq:xxx_b_k}
   b_k=\frac{1}{2j}\bigg(\frac{k}{2j+1}+\frac{j-k}{1+j(1+w)}\bigg)
\end{equation}
and
\begin{equation}
   \label{eq:xxx_u_k}
   u_k=\frac{(w-1)\sqrt{k(2j+1-k)}}{2(2j+1)[1+j(1+w)]}.
\end{equation}
The eigenvalues (statistical weights) of matrix (\ref{eq:rhoGxxx}) are equal to
\begin{eqnarray}
   \label{eq:xxx_pq}
   &&p_k=p_0=p_{4j+1}
	 =\frac{1}{2[1+j(1+w)]},
   \nonumber\\
   &&q_k
   =\frac{w}{2[1+j(1+w)]}
\end{eqnarray}
with $k=1,2,\ldots,2j$.

Using the above expressions and equations from Appendix~\ref{sec:append},
Eqs.~(\ref{eq:F0})--(\ref{eq:U0U1}), we find that the  0- and 1-branches of XXX model
coincide (${\cal F}={\cal F}_1={\cal F}_0$ and ${\cal U}={\cal U}_1={\cal U}_0$), and
as a result, quantum correlations LQFI and LQU are given by
\begin{equation}
   \label{eq:Fxxx}
   {\cal F}(T,j)=\frac{4j(j+1)(1-w)^2}{3(2j+1)(1+j+jw)(1+w)}
\end{equation}
and
\begin{equation}
   \label{eq:Uxxx}
   {\cal U}(T,j)=\frac{4j(j+1)(\sqrt{w}-1)^2}{3(2j+1)[1+j(w+1)]},
\end{equation}
respectively.

So, the formulas for quantum correlations have been obtained and now we can move on to
studying their properties.

\subsubsection{LQU and LQFI vs temperature}
\label{subsec:T-behav}
\begin{figure}[t]
\begin{center}
\epsfig{file=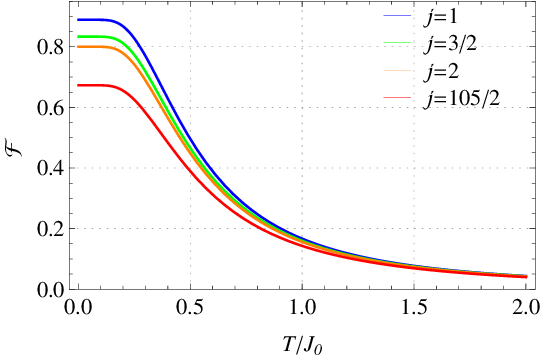,width=7.8cm}
		\put(-185,150){($\rm\bf a$)}
\hspace{3mm}
\epsfig{file=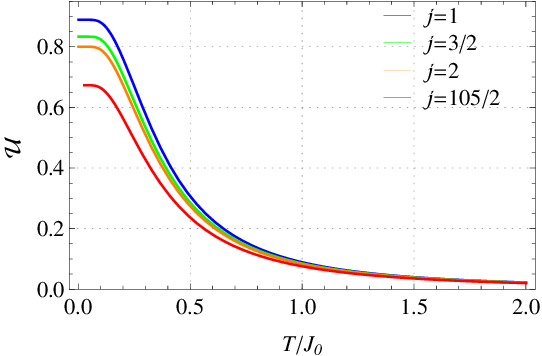,width=7.8cm}
		\put(-185,150){($\rm\bf b$)}
\end{center}
\begin{center}
\caption{(Color online)
Local quantum Fisher information $\cal F$ ($\rm\bf a$) and local quantum uncertainty
$\cal U$ ($\rm\bf b$) versus reduced temperature $T/J_0$ for the antiferromagnetic
Heisenberg systems $(j,1/2)$ with $j=1$ (blue), $j=3/2$ (green), $j=2$ (orange) and
$j=105/2$ (red).
}
\label{fig:zXXX_af}
\end{center}
\end{figure}

\begin{figure}[t]
\begin{center}
\epsfig{file=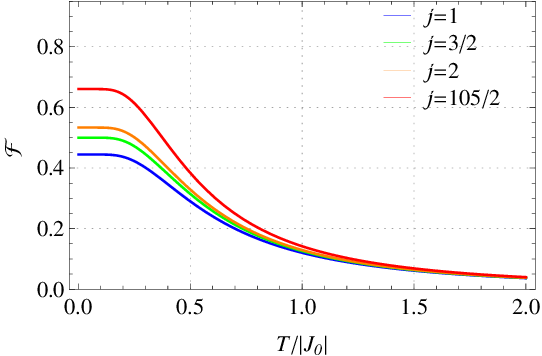,width=8.4cm}
		\put(-185,145){($\rm\bf a$)}
\hspace{3mm}
\epsfig{file=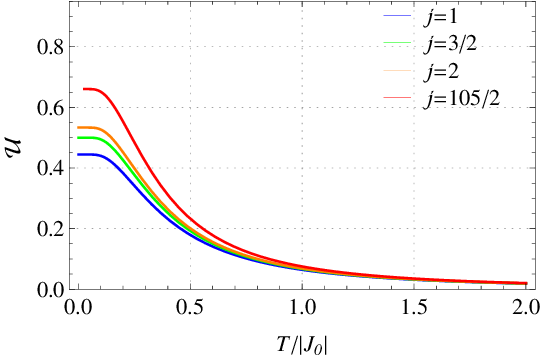,width=8.4cm}
		\put(-185,145){($\rm\bf b$)}
\end{center}
\begin{center}
\caption{(Color online)
Local quantum Fisher information $\cal F$ ($\rm\bf a$) and local quantum uncertainty
$\cal U$ ($\rm\bf b$) versus reduced temperature $T/|J_0|$ for the ferromagnetic
Heisenberg systems $(j,1/2)$ with $j=1$ (blue), $j=3/2$ (green), $j=2$ (orange) and
$j=105/2$ (red).
}
\label{fig:zXXX_f}
\end{center}
\end{figure}

Let us first discuss the behavior of quantum correlations against temperature.
Figures~\ref{fig:zXXX_af} and \ref{fig:zXXX_f} show the corresponding dependences of
the quantum correlations (\ref{eq:Fxxx}) and (\ref{eq:Uxxx}) for antiferromagnetic
($J_0>0$) and ferromagnetic ($J_0<0$) couplings, respectively.
%
As the temperature increases, the correlations first maintain a quasi-stationary value
and then monotonically tend to zero.
Using Eqs.~(\ref{eq:Fxxx}), (\ref{eq:Uxxx}) and (\ref{eq:w}), we find that at high
temperatures, the correlations vary as
\begin{equation}
   \label{eq:F_highT_j}
   {\cal F}(T,j)\approx\frac{1}{6}\Big(\frac{J_0^2}{T^2}+\frac{J_0^3}{4T^3\sqrt{j(j+1)}} +\dots\Big),\qquad
\end{equation}
\begin{equation}
   \label{eq:U_highT_j}
   {\cal U}(T,j)\approx\frac{1}{12}\Big(\frac{J_0^2}{T^2}+\frac{J_0^3}{4T^3\sqrt{j(j+1)}} +\dots\Big).
\end{equation}
Thus, both correlations fall to zero according to the power laws.
The main asymptotic terms behave as $\sim T^{-2}$ and are independent of $j$ and the sign of
$J_0$; additionally, $\cal F$ is twice as large as $\cal U$.
Note, for comparison, that the ordinary spin-spin correlation function
$G(T)=\langle{\bf S}\cdot{\bf s}\rangle$ varies as $1/T$ at high temperatures.

As seen from Fig.~\ref{fig:zXXX_af}, the values of quantum correlations at zero
temperature decrease with increasing spin number $j$.
This is in agreement with intuitive expectations: as $j\to\infty$, the particle $A$
becomes more and more classical, and therefore, the quantum correlations gradually
weaken.

The case of ferromagnetic interactions ($J_0<0$) is illustrated in
Fig.~\ref{fig:zXXX_f}a,b.
In a high-temperature region, both quantum correlations LQFI and LQU fall to zero
according to the same power law $T^{-2}$.

However, the situation in the lower-temperature region is strange.
Indeed, as can be seen from Fig.~\ref{fig:zXXX_f}, both discord-type quantum
correlations {\em increase} with increasing the magnitude of spin $j$.
This contradicts our expectations, according to which the normalized components of the
spin operators $\bf S$ transform into the components $r_x$, $r_y$ and $r_z$ of a
classical vector $\bf r$
on a sphere of unit radius and the Hamiltonian
(\ref{eq:Hxxx}) is reduced to the ${\cal H}=J_0{\bf r}\cdot{\bf s}$, that is, to the
one-qubit system, where all quantum correlations are absent.
We will return to the discussion of this issue in Subsec.~\ref{subsec:T0GibbsXXX}.
%
%

Remarkably, the case of the isotropic Heisenberg model (\ref{eq:Hxxx}) satisfies a much
higher symmetry than axially symmetric systems.
Such a system has rotational symmetry SU(2), and its states commute with all components
of the total spin (not just with its $z$-component).
The SU(2)-invariant density matrix in the total spin basis is given by \cite{S03,S05}
\begin{eqnarray}
   \label{eq:rhoSU2}
   \rho&=&\frac{F}{2j}\sum_{m=-j+1/2}^{j-1/2}|j-1/2,m\rangle\langle j-1/2,m|
   \nonumber\\
    &+&\frac{1-F}{2(j+1)}\sum_{m=-j-1/2}^{j+1/2}|j+1/2,m\rangle\langle j+1/2,m|,\
\end{eqnarray}
where the single parameter $F\in[0,1]$.
The quantity $F/2j$ is the $2j$-fold degenerated eigenvalue $q_k$,
Eq.~(\ref{eq:xxx_pq}), of the density matrix (\ref{eq:rhoGxxx}).
Thus, the single parameter $F$ is derived as
\begin{eqnarray}
   \label{eq:F}
   F&=&\frac{jw}{1+j+jw}
   \nonumber\\
   &=&1-\frac{j+1}{1+j+j\exp{[(2j+1)J_0/(2T\sqrt{j(j+1)})}]}.
   \nonumber\\
\end{eqnarray}

For the SU(2)-invariant model, some remarkable results have been obtained
\cite{S03,S05,MC08,WW08,CG13,WMFW14,FA15,WGFW22}.
In particular, it was found that quantum entanglement, measured by (double) negativity,
is given as \cite{S03,S05,MC08}
\begin{equation}
   \label{eq:NN_F}
   {\cal N}=\max{\Big[0,2\Big(F-\frac{2j}{2j+1}\Big)\Big]},
\end{equation}
and the entropic entanglement of formation (EoF) is equal to \cite{MC08}
\begin{equation}
   \label{eq:EoF_F}
   {\rm EoF}=
   \begin{cases}
   0, & F\le\frac{2j}{2j+1}\cr
   h\big(\frac{1}{2j+1}\big(\sqrt{F}-\sqrt{2j(1-F)}\big)^2\big), & F>\frac{2j}{2j+1},
   \end{cases}
\end{equation}
where $h(x)=-x\log_2{x}-(1-x)\log_2{(1-2)}$ is the binary entropy.
Substituting the expression for $F$, Eq.~(\ref{eq:F}), yields negativity as functions
of temperature and spin $j$,
\begin{equation}
   \label{eq:NN_T}
   {\cal N}(T,j)=\frac{4je^{-jJ_0/2T\sqrt{j(j+1)}}}{(2j+1)Z}\max{\big[0,w-2(j+1)\big]}.
\end{equation}
Similarly, using the expression for $F$, we arrive at EoF$(T,j)$.

Next, the QD is given by \cite{CG13}
\begin{equation}
   \label{eq:QD}
   Q=1+F\log_2{\frac{F}{2j}}+(1-F)\log_2{\frac{1-F}{2j+2}}
   -\sum_{n=0}^{\lfloor j\rfloor}\lambda_n^{\pm}\log_2{\lambda_n^{\pm}},
\end{equation}
where
\begin{equation}
   \label{eq:lam}
   \lambda_n^{\pm}=\frac{1}{2j+1}\pm\frac{j-n}{j(j+1)(2j+1)}|(2j+1)F-j|
\end{equation}
and $\lfloor.\rfloor$ denotes the floor function.
Again taking the expression for the parameter $F$ (\ref{eq:F}), Eqs.~(\ref{eq:QD}) and
(\ref{eq:lam}) allow us to get the function $Q(T,j)$.

Further, the LQU of SU(2)-invariant states was derived to be
\cite{FA15}
\begin{equation}
   \label{eq:LQU_j}
   {\cal U}=\frac{8j(j+1)}{3(2j+1)}\Bigg(\sqrt{\frac{F}{2j}}-\sqrt{\frac{1-F}{2(j+1)}}\Bigg)^2.
\end{equation}
Our result (\ref{eq:Uxxx}) is in full agreement with this formula.

\begin{figure}[t]
\begin{center}
\epsfig{file=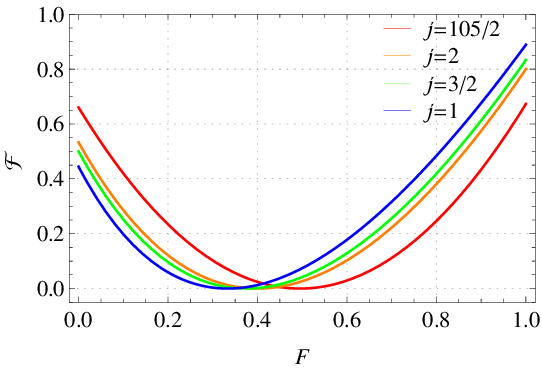,width=8.4cm}
\caption{(Color online)
Local quantum Fisher information as a function of $F$ for the different spin angular
momentum quantum numbers $j=1$ (blue), $j=3/2$ (green), $j=2$ (orange) and $j=105/2$
(red).
Points $F=0$ and $F=1$ correspond to the zero temperature $T=0$ for a ferromagnet and
an antiferromagnet, respectively.
The local minima at $F=j/(2j+1)$ correspond to the infinitely high temperatures.
}
\label{fig:zXXX_LQFIvsF}
\end{center}
\end{figure}

Using Eqs.~(\ref{eq:Fxxx}) and (\ref{eq:F}), we express LQFI through $F$ and $j$
parameters,
\begin{equation}
   \label{eq:Fxxx-vsF}
   {\cal F}=\frac{4}{3}\frac{(j-(2j+1)F)^2}{(2j+1)(j+F)}.
\end{equation}
Figure~\ref{fig:zXXX_LQFIvsF} shows this function versus $F$.
%
Both $\cal F$ and its first derivative vanish at $F=F_c=j/(2j+1)$, exactly at the same
value of $F$ as in the QD and LQU cases, and half as much as for EoF and negativity;
see \cite{MC08,CG13,FA15}.
It is also clear from Fig.~\ref{fig:zXXX_LQFIvsF} that LQFI decreases with increasing
spin $j$ in the region $F>F_c$, and, conversely, increases in the region $F<F_c$ as
$j\to\infty$.

Thus, the use of SU(2) invariance has yielded many interesting quantum correlations for
the fully isotropic two-site Heisenberg model, and we will apply them below.


\subsubsection{Threshold temperature}
\label{subsubsec:T_th}
As is known, quantum entanglement, unlike other types of quantum correlations, can
suddenly disappear in the process of evolution with respect to some parameters (time,
temperature, etc.).
This unusual phenomenon is called entanglement sudden death (ESD) \cite{YE09}.

\begin{figure}[t]
\begin{center}
\epsfig{file=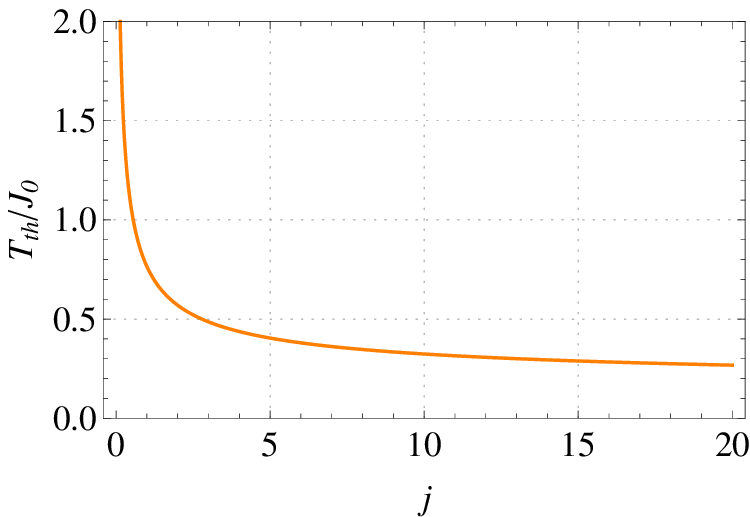,width=8.4cm}
\end{center}
\begin{center}
\caption{(Color online)
Normalized threshold temperature $T_{\rm th}/J_0$ as a function of spin parameter $j$.
}
\label{fig:zCom}
\end{center}
\end{figure}

Both EoF (\ref{eq:EoF_F}) and negativity (\ref{eq:NN_T}) initially change with
increasing temperature, similar to the discord-like measures depicted in
Fig.~\ref{fig:zXXX_af}.
Then, as follows from Eqs.~(\ref{eq:EoF_F}) and (\ref{eq:NN_T}), they vanish at the
threshold temperature
\begin{equation}
   \label{eq:T_th}
   T_{\rm th}=\frac{(2j+1)J_0}{2\sqrt{j(j+1)}\ln{(2j+2)}}
\end{equation}
and continue to be absent above it.
From here we come to an important conclusion:
\begin{equation}
   \label{eq:T_th8}
   \lim_{j\to\infty}T_{\rm th}=0.
\end{equation}
Thus, the threshold temperature monotonically (logarithmically slowly) tends to zero
with increasing spin and disappears in the limit of infinitely large $j$.
The graph of $T_{\rm th}/J_0$ versus $j$ is shown in Fig.~\ref{fig:zCom}.
The results obtained are consistent with intuitive expectations based on physical
reasoning.
%

It is worth noting the following.
If the spin operator ${\bf S}$ is not normalized to its length
$|{\bf S}|=\sqrt{j(j+1)}\approx j$, then instead of (\ref{eq:T_th}), the threshold
temperature will be equal to $T_{\rm th}=(2j+1)J_0/2\ln{(2j+2)}$, which leads to the
result opposite to (\ref{eq:T_th8}), namely
$\lim_{j\to\infty}T_{\rm th}=\infty$ \cite{WW06}.

The question of whether quantum correlations increase with increasing spin or not could
be decided by an experiment, which is a measure of truth.
The entanglement temperature (i.e., threshold temperature) for various molecular
cluster magnets and rare-earth ions was estimated in \cite{PSS09,DCSR13}.
Using entanglement witnesses built from measurements of magnetic susceptibility and
mean energy, the authors found an increase in the threshold temperature with increasing
spin and orbital angular momenta.

Unfortunately, the used witnesses directly depend on whether or not the spin
normalization is taken into account.
The normalization of operators was not performed by the authors \cite{PSS09,DCSR13}.
It is easy to check that if the spin and angular momenta are rescaled according to the
correspondence principle, then the same experimental data will show a decrease in
entanglement temperatures.
Thus, such an experiment cannot resolve the above question.

Finally, many authors \cite{SW07,HWSY08,LRKL12,LLWYC14,CS20,VS21,S21,VST22} continue to
claim to this day that the magnitude of spins may supposedly enlarge the threshold
temperature and thereby enhance thermal entanglement at sufficiently high (room)
temperatures.
The reason for these misconceptions is still the same: ignoring the normalization of
spin operators and, as a consequence, not fulfilling the correspondence principle.


\subsubsection{Ground-state quantum correlations}
\label{subsec:T0GibbsXXX}
Let us discuss the limit $T\to0$ of the XXX system (\ref{eq:Hxxx}) in more detail.
If $J_0>0$ (antiferromagnetic coupling), the big and small spins form an antiparallel
configuration $(\Uparrow\downarrow)$, and the spins will be parallel to each other,
$(\Uparrow\uparrow)$, when the interactions are ferromagnetic ($J_0<0$).

As noted earlier, quantum correlations exhibit temperature-independent, stationary
behavior in the low-temperature region (see again Figs.~\ref{fig:zXXX_af} and
\ref{fig:zXXX_f}).
Therefore, our results will also be valid in some vicinity above absolute zero
temperature.

At $T=0$, the entries of the density matrix (\ref{eq:rhoGxxx}) and its eigenvalues are
significantly simplified and reduced to
\begin{equation}
   \label{eq:xxx_ak_T0}
   a_k=
   \begin{cases}
   (2j+1-k)/[2j(2j+1)], &{\rm if}\ J_0>0\cr
   k/[2(j+1)(2j+1)], &{\rm if}\ J_0<0,
   \end{cases}
\end{equation}
\begin{equation}
   \label{eq:xxx_a2sk_T0_}
   b_k=
   \begin{cases}
   k/[2j(2j+1)], &{\rm if}\ J_0>0\cr
   (2j+1-k)/[2(j+1)(2j+1)], &{\rm if}\ J_0<0,
   \end{cases}
\end{equation}
\begin{equation}
   \label{eq:xxx_uk_T0}
   u_k=
   \begin{cases}
   -\sqrt{k(2j+1-k)}/[2j(2j+1)], &{\rm if}\ J_0>0\cr
   -\sqrt{k(2j+1-k)}/[2(j+1)(2j+1)], &{\rm if}\ J_0<0,
   \end{cases}
\end{equation}
\begin{equation}
   \label{eq:xxx_p0_p2s1_T0}
   p_k=p_0=p_{4j+1}=
   \begin{cases}
   0, &{\rm if}\ J_0>0\cr
   1/[2(j+1)], &{\rm if}\ J_0<0,
   \end{cases}
\end{equation}
and
\begin{equation}
   \label{eq:xxx_qk_T0}
   q_k=
   \begin{cases}
   1/2j, &{\rm if}\ J_0>0\cr
   0, &{\rm if}\ J_0<0.
   \end{cases}
\end{equation}
The $F$ parameter (\ref{eq:F}) is now given by
\begin{equation}
   \label{eq:xxx_F_T0}
   F(T=0,j)=
   \begin{cases}
   1, &{\rm if}\ J_0>0\cr
   0, &{\rm if}\ J_0<0.
   \end{cases}
\end{equation}
Using these expressions and Eqs.~(\ref{eq:Fxxx}) and (\ref{eq:Uxxx}), we find that
\begin{equation}
   \label{eq:xxx_F0_T0}
   {\cal F}(0,j)={\cal U}(0,j)
   =\frac{2}{3}\bigg(1+\frac{1}{2j+1}\frac{J_0}{|J_0|}\bigg).
\end{equation}
Importantly, the magnitude of coupling $J_0$ together with spin normalization
coefficients have dropped out; only the sign of $J_0$ is essential in the
low-temperature region.
In addition, the zero-temperature quantum correlations LQFI and LQU coincide with each
other, and therefore, we will denote them as ${\cal F}/{\cal U}$ or LQFI/LQU.
Selected values of this single correlation are collected in Table~\ref{tab:UFat0} for
reference.
Finally, in the limit $j\to\infty$, ${\cal F}={\cal U}=2/3=0.6666\ldots$.
%
\begin{table}[t]
\caption{\label{tab:UFat0}%
Ground-state quantum correlation LQFI/LQU in the spin-$(j,1/2)$ XXX system with both
antiferromagnetic ($J_0>0$) and ferromagnetic ($J_0<0$) interactions.
}
\begin{ruledtabular}
\begin{tabular}{cll}
\textrm{$j$}&
\textrm{$J_0>0$}&
\multicolumn{1}{c}{\textrm{$J_0<0$}}\\
\noalign{\smallskip}\hline\noalign{\smallskip}
1/2 & 1 & 1/3$=0.33(3)$\\
1 & 8/9=0.88(8) &  4/9=0.44(4)  \\
3/2 & 5/6=0.83(3) &  1/2=0.5  \\
2 & 4/5=0.8 &  8/15=0.53(3)  \\
105/2 & 0.67295 &  0.66037  \\
\end{tabular}
\end{ruledtabular}
\end{table}
%

In accord with Eqs.~(\ref{eq:NN_F}), (\ref{eq:EoF_F}) and (\ref{eq:xxx_F_T0}), the EoF
at zero temperature equals
\begin{equation}
   \label{eq:xxx_EoF_T0}
   {\rm EoF}(0,j)=h\big(1/(2j+1)\big)
\end{equation}
and the double negativity is
\begin{equation}
   \label{eq:xxx_Neg_T0}
   {\cal N}(0,j)=\frac{2}{2j+1}.
\end{equation}

%
\begin{figure}[t]
\begin{center}
\epsfig{file=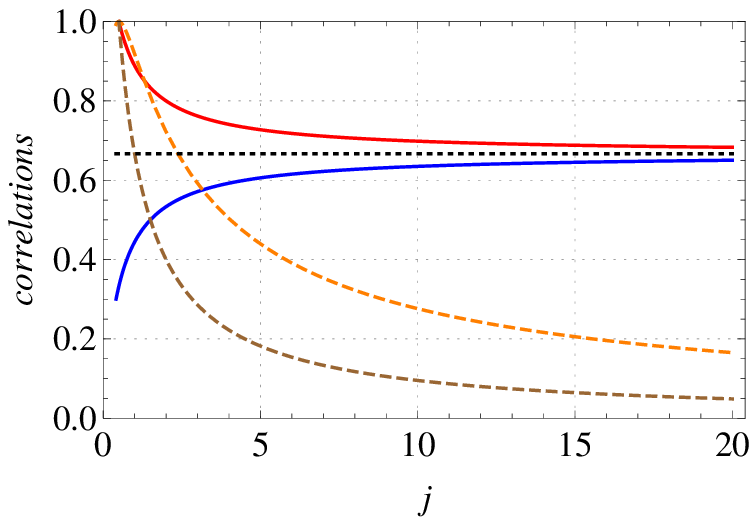,width=8.4cm}
\end{center}
\begin{center}
\caption{(Color online)
Ground-state quantum correlations vs $j$ in the XXX system: LQFI/LQU for $J_0>0$ (red
line), LQFI/LQU for $J_0<0$ (blue line), black dashed horizontal line corresponds to
their asymptotic value 2/3, EoF (orange dashed line) and double negativity
$\cal N$ (brown dashed line).
}
\label{fig:zQCxxx_T0}
\end{center}
\end{figure}
%
\begin{figure}[t]
\begin{center}
\epsfig{file=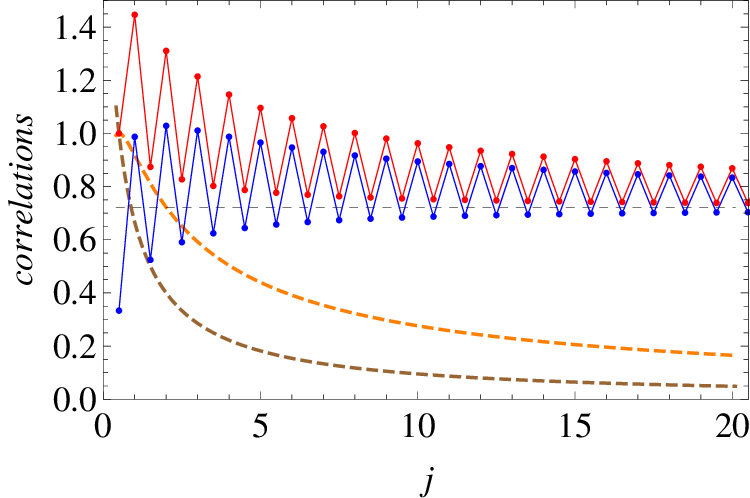,width=8.4cm}
\end{center}
\begin{center}
\caption{(Color online)
Ground-state quantum correlations vs $j$ in the XXX system.
QD (red dots) for $J_0>0$ and QD (blue dots) for $J_0<0$, connected by red and blue
broken lines, respectively; black dashed horizontal line corresponds to their
asymptotic value $0.7213$, EoF (orange dashed line) and double negativity
$\cal N$ (brown dashed line).
}
\label{fig:zQCxxx_T0_}
\end{center}
\end{figure}
%
Figure~\ref{fig:zQCxxx_T0} shows the behavior of different ground-state quantum
correlations in the pure isotropic Heisenberg model whose state is SU(2) invariant.
Let us first consider the quantum entanglement.
It exists only in the antiferromagnetic system ($J_0>0$) and is completely absent in
the case of ferromagnetic interactions ($J_0<0$).
Quantum entanglement measured by EoF and double negativity equals unity at $j=1/2$.

As the spin increases, both forms of entanglement monotonically degrade and completely
disappear in the limit of an infinitely large value of $j$.
This is in full agreement with intuitive expectations.
Indeed, when the normalized spin operators become almost commuting quantities, the
system approaches more and more the classical limit and the quantum correlations
weaken.

The picture is quite different for the $\cal F/U$ (LQFI/LQU) correlation.
Look again at Fig.~\ref{fig:zQCxxx_T0}.
For an antiferromagnetic coupling ($J_0>0$), LQU/LQFI (red solid line) is equal to one
at $j=1/2$ and initially begins to fall with increasing $j$ (see also
Table~\ref{tab:UFat0}).
However, this correlation then asymptotically tends to the stationary level 2/3, rather
than to zero.
For $J_0<0$ (ferromagnetic coupling), the discord-like quantum correlation LQU/LQFI
(blue solid line) is one third for $j=1/2$, and then starts {\em increasing} with
increasing $j$ to the same mysterious value 2/3.

The QD (\ref{eq:QD}) shows more complex behavior, see Fig.~\ref{fig:zQCxxx_T0_}.
It jumps between integer and half-integer values of spin $j$.
The amplitude of such jumps (oscillations) decreases with increasing $j$ and the
QD asymptotically again tends to a non-zero residual value (now equal to
$0.7213\ldots$) in the limit of infinitely large spin.
%



\subsection{Heisenberg XXZ model}
\label{subsec:XXZ}
Consider now the anisotropic model with the Hamiltonian
\begin{equation}
   \label{eq:Hxxz}
   {\cal H}=J_zS_z\otimes\sigma_z+J(S_x\otimes\sigma_x+S_y\otimes\sigma_y).
\end{equation}
The SU(2) symmetry is broken in this case.
Spin normalizations are omitted here since we are only interested in ground-state
correlations.

The Hamiltonian (\ref{eq:Hxxz}) in open matrix form is
\begin{widetext}
\begin{equation}
   \label{eq:Hxxz_open}
   {\cal H}=
	 \left(
      \begin{array}{cccccccccccc}
      jJ_z&\ &\ &\ &\ &\ &\ &\ &\ &\ &\ &\ \\
      \ &-jJ_z&J\sqrt{2j}&\ &\ &\ &\ &\ &\ &\ &\ &\ \\
      \ &J\sqrt{2j}&(j-1)J_z&\ &\ &\ &\ &\ &\ &\ \ &\ &\ \\
      \ &\ &\ &\ddots&\ &\ &\ &\ &\ &\ &\ &\ \\
      \ &\ &\ &\ &(-j+k-1)J_z&J\sqrt{k(2j-k+1)}&\ &\ &\ &\ &\ &\ \\
      \ &\ &\ &\ &J\sqrt{k(2j-k+1)}&(j-k)J_z&\ &\ &\ &\ &\ &\ \\
      \ &\ &\ &\ &\ &\ &\ &\ddots&\ &\ &\ &\ \\
      \ &\ &\ &\ &\ &\ &\ &\ &\ &(j-1)J_z&J\sqrt{2j}&\ \\
      \ &\ &\ &\ &\ &\ &\ &\ &\ &J\sqrt{2j}&-jJ_z&\ \\
			\ &\ &\ &\ &\ &\ &\ &\ &\ &\ &\ &jJ_z
      \end{array}
   \right).
\end{equation}
\end{widetext}
This matrix preserves the axially symmetrical structure (\ref{eq:Has}).

Using the general expressions from Sec.~\ref{sec:ASH}, we find that the energy spectrum
of the system consists of two levels $E_0=jJ_z$ and $2j$ pairs with energies
\begin{equation}
   \label{eq:E12k}
   E_k^{(1,2)}=\frac{1}{2}(-J_z\pm R_k),
\end{equation}
where
\begin{equation}
   \label{eq:Rk}
   R_k=\sqrt{(2k-2j-1)^2J_z^2+4k(2j-k+1)J^2}
\end{equation}
and $k=1,\ldots,2j$.
The partition function is given as
\begin{equation}
   \label{eq:Zxxz}
   Z=2e^{-jJ_z/2T}+2e^{J_z/2T}\sum_{k=1}^{2j}\cosh{(R_k/2T)}.
\end{equation}
It is even with respect to the coupling constant $J$.
The statistical weights are equal to
\begin{eqnarray}
   \label{eq:Pxxz}
   &&p_0=p_{4j+1}=\frac{1}{Z}e^{-jJ_z/T},
   \nonumber\\
   &&p_k=\frac{1}{Z}e^{(J_z-R_k)/2T},\\
   &&q_k=\frac{1}{Z}e^{(J_z+R_k)/2T}.
   \nonumber
\end{eqnarray}
\begin{figure}[b]
\begin{center}
\epsfig{file=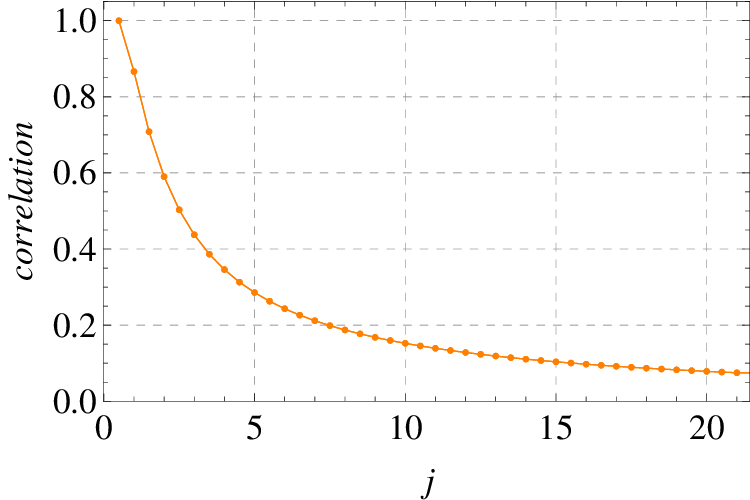,width=8.4cm}
\end{center}
\begin{center}
\caption{(Color online)
Ground-state quantum correlation ${\cal F}/{\cal U}$ vs $j$ in the XXZ model with
interactions $J_z=1$ and $J=0.9$.
}
\label{fig:zXXZ_FUp109}
\end{center}
\end{figure}
\begin{figure}[t]
\epsfig{file=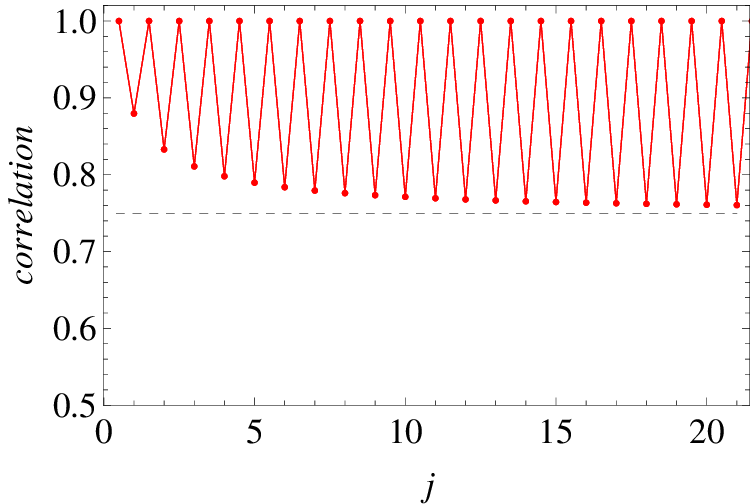,width=8.4cm}
		\put(-185,50){($\rm\bf a$)}
\hspace{3mm}
\epsfig{file=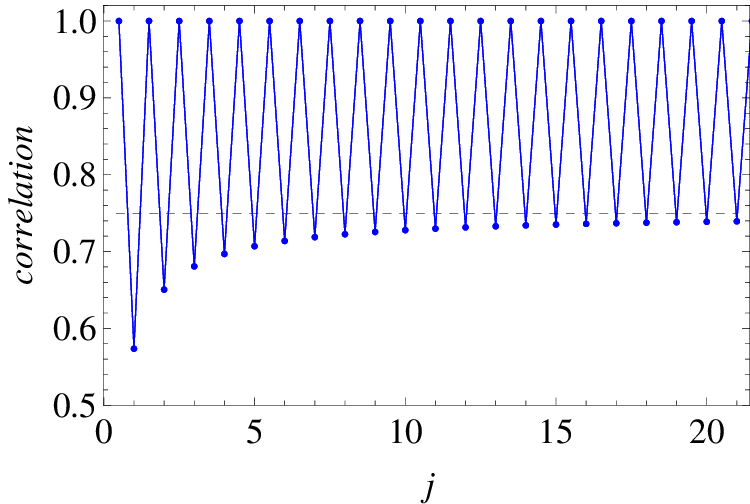,width=8.4cm}
		\put(-185,50){($\rm\bf b$)}
\caption{(Color online)
Ground-state quantum correlation ${\cal F}/{\cal U}$ vs $j$ in the XXZ system.
({\bf a}) red dots (connected for clarity by a red broken line)
represent the quantum correlation ${\cal U}_1={\cal F}_1$ for $J_z=1$ and $J=1.1$.
({\bf b}) blue dots and dotted line are the same as in the previous case ({\bf a}), but
for $J_z=-1$ and $J=1.1$.
Black dotted horizontal line in both panels corresponds to the level $0.7499\ldots$.
}
\label{fig:zXXZ_FU111}
\end{figure}
%
Finally, the matrix elements of the Gibbs density matrix (\ref{eq:rho}) are expressed
as follows
\begin{eqnarray}
   \label{eq:abu_xxz}
   &&a_k=\frac{1}{Z}\Big(\cosh{\frac{R_k}{2T}}+J_z\frac{2(j-k)+1}{R_k}\sinh{\frac{R_k}{2T}}\Big)e^{J_z/2T},
   \nonumber\\
   &&b_k=\frac{1}{Z}\Big(\cosh{\frac{R_k}{2T}}-J_z\frac{2(j-k)+1}{R_k}\sinh{\frac{R_k}{2T}}\Big)e^{J_z/2T},
   \nonumber\\
   &&u_k=-\frac{2J\sqrt{k(2j-k+1)}}{ZR_k}\sinh{\frac{R_k}{2T}}e^{J_z/2T}.
\end{eqnarray}
Note that $a_k$, $b_k$, and $u_k^2$ are even with respect to $J$ for
$\forall k=1,\ldots,2j$.
Consequently, all four branches of quantum correlations LQU and LQFI, determined by
Eqs.~(\ref{eq:F0})-(\ref{eq:F1a}), and hence the quantum correlations $\cal F$ and
$\cal U$ themselves, respectively Eqs.~(\ref{eq:F0F1}) and (\ref{eq:U0U1}), are also
invariant with respect to the coupling constant $J$.

If $J=0$, the system (\ref{eq:Hxxz}) becomes Ising, where there are no quantum
correlations.
Thus, it makes sense to consider only $J\neq0$.
We will focus on the system in the ground state and try to interpret the anomalies that
have been observed in the behavior of discord-type correlations.

Let us consider the system with $J<|J_z|$.
For any small weakening of the transverse interaction $J$, the behavior of discord-like
quantum correlations shown in Fig.~\ref{fig:zQCxxx_T0} changes dramatically.
Indeed, the former ferromagnetic correlation (blue line) completely disappears, the 2/3
level (dashed black horizontal line) vanishes, and the ex-antiferromagnetic correlation
(red line) now asymptotically tends to zero with increasing spin $j$.
The new behavior of quantum correlation is shown in Fig.~\ref{fig:zXXZ_FUp109}.

Note that the correlation ${\cal F}={\cal U}$ here is determined by the 0-branches,
which are given by (\ref{eq:F0}) and (\ref{eq:U0}).

The behavior of quantum correlation is now entirely consistent with expectations from a
physical point of view.
Thus, the reason for the strange dependence of discord-type correlations on spin in the
XXX system at zero temperature is rooted in {\em instability}.
The SU(2) state is too symmetric, and even a small perturbation changes the behavior of
quantum correlations based on (optimal) measurements.
Although the correlations are unstable, the SU(2) state can nevertheless be interpreted
as an example of a state that preserves quantumness of correlations in the classical
limit.

However, new difficulties arise when $J>|J_z|$.
With increasing strength of transverse coupling $J$, the system moves away from the
Ising limit, becoming more quantum.
The behavior of quantum correlations in this case is shown in
Fig.~\ref{fig:zXXZ_FU111}a,b.

The quantum correlations here are determined by the 1-branches, which are defined by
Eqs.~(\ref{eq:U1})--(\ref{eq:F1}).

As for QD (see Fig.~\ref{fig:zQCxxx_T0_}), quantum correlations for
both ferromagnetic and antiferromagnetic couplings $J_z$ oscillate between integer and
half-odd-integer spins $j$.
The values of these correlations for half-integer spins are equal to one.
The values of antiferromagnetic correlation (red dots in Fig.~\ref{fig:zXXZ_FU111}a)
tend to the level of 0.7499 from above with increasing magnitudes of integer spin.
On the other hand, the values of correlation in the ferromagnetic case (blue dots in
Fig.~\ref{fig:zXXZ_FU111}b) tend to the same level from below again for integer spins.
The amplitude of the jumps (oscillations) here does not vanish with increasing $j$.


%

\subsection{Heisenberg XXZ model in non-uniform field}
\label{subsec:XXXB1B2}
Let us move on to the consideration of the anisotropic Heisenberg XXZ system, subjected
to external inhomogeneous magnetic fields applied along the $z$-axis.
The Hamiltonian is given as
\begin{equation}
   \label{eq:HxxzB1B2}
   {\cal H}=J_zS_z\otimes\sigma_z+J(S_x\otimes\sigma_x+S_y\otimes\sigma_y)+B_1S_z+B_2\sigma_z.
\end{equation}
External magnetic fields $B_1$ and $B_2$ destroy the Z$_2$ spin reversal symmetry that
is present in both XXX (\ref{eq:Hxxx}) and XXZ (\ref{eq:Hxxz}) systems.
However, the axial symmetry invariance is preserved, and the Hamiltonian
(\ref{eq:HxxzB1B2}) in open matrix form looks like
\begin{widetext}
\begin{equation}
   \label{eq:HxxzB1B2open}
   {\cal H}=
	 \left(
      \begin{array}{cccccccc}
      j(J_z+B_1)+B_2&\ &\ &\ &\ &\ &\ &\ \\
      \ &\ddots&\ &\ &\ &\ &\ &\ \\
      \ &\ &(-j+k-1)(J_z-B_1)-B_2&J\sqrt{k(2j-k+1)}&\ &\ &\ &\ \\
      \ &\ &J\sqrt{k(2j-k+1)}&(j-k)(J_z+B_1)+B_2&\ &\ &\ &\ \\
      \ &\ &\ &\ &\ &\ddots&\ &\ \\
			\  &\ &\ &\ &\ &\ &\ &j(J_z-B_1)-B_2
      \end{array}
   \right).
\end{equation}
\end{widetext}
Comparing this matrix with (\ref{eq:Has}), we establish that
\begin{eqnarray}
   \label{eq:Hij_}
   &&E_0=j(J_z+B_1)+B_2,
   \nonumber\\
   &&E_{4j+1}=j(J_z-B_1)-B_2,
   \nonumber\\
   &&h_k=(-j+k-1)(J_z-B_1)-B_2,\\
   &&h^\prime_k=(j-k)(J_z+B_1)+B_2,
   \nonumber\\
   &&g_k=J\sqrt{k(2j-k+1)}.
   \nonumber
\end{eqnarray}
Using now the general formulas of Sec.~\ref{sec:ASH}, we find the required density
matrix, which allows us to calculate the quantum correlations given in
Appendix~\ref{sec:append}.

\begin{figure}[t]
\begin{center}
\epsfig{file=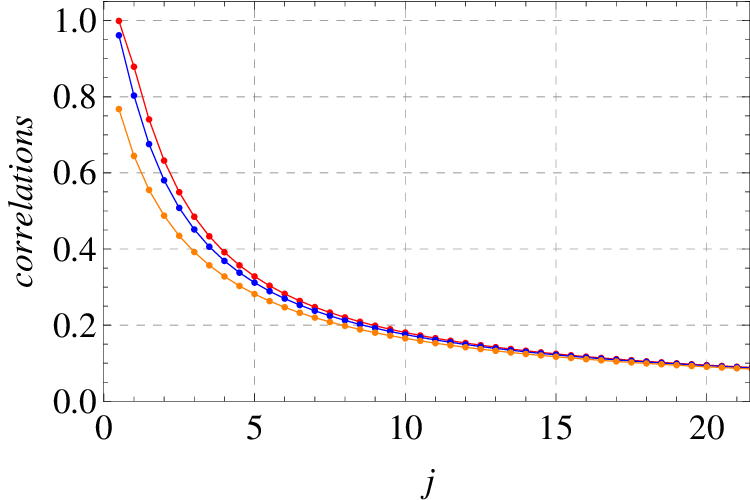,width=8.4cm}
\end{center}
\begin{center}
\caption{(Color online)
Ground-state quantum correlation LQFI/LQU vs $j$ in the antiferromagnetic XXX system
($J_z=J=1$) with external fields:
$B_1=0.05$, $B_2=0$ (red); $B_1=0$, $B_2=0.2$ (blue); $B_1=-0.5$, $B_2=0.3$ (orange).
}
\label{fig:zQCxxx_data123}
\end{center}
\end{figure}

The results are as follows.
Let us consider again the isotropic XXX system, but now perturbed by an external field,
see Fig.~\ref{fig:zQCxxx_data123}.
%
%
First of all, it can be stated that external magnetic fields of any intensity radically
change, similar to the case of the XXZ system, the behavior of discord-type
correlations.
This provides new evidence for the instability of these quantum correlations in the
SU(2) quantum state.

\begin{figure}[t]
\begin{center}
\epsfig{file=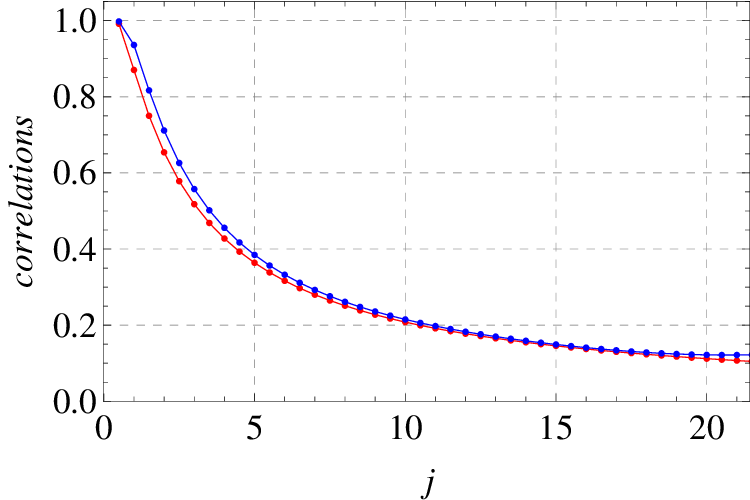,width=8.4cm}
\end{center}
\begin{center}
\caption{(Color online)
Ground-state quantum correlations LQFI/LQU vs $j$ in the XXZ system with:
$J_z=1$, $J=1.1$, $B_1=0.2$, $B_2=0$ (red) and
$J_z=-1$, $J=1.1$, $B_1=0.2$, $B_2=0$ (blue).
}
\label{fig:zQCxxz_data45}
\end{center}
\end{figure}

Turn now to the oscillations (variations) of quantum correlations depicted in
Fig.~\ref{fig:zXXZ_FU111}.
Figure~\ref{fig:zQCxxz_data45} clearly demonstrates that applying a magnetic field
eliminates all jumps in the correlations, and, in addition,
the correlations now decay monotonically to zero with increasing spin $j$.
%
%
This behavior again restores a physically acceptable picture.

Unfortunately, there is no solution for the entropic QD ($Q$) taking into
account the external field.
Nevertheless, based on the above, one may assume that the external field will also
eliminate its oscillations, shown in Fig.~\ref{fig:zQCxxx_T0_}, and everything will
fall into place again.
Jumps are a manifestation of instabilities of quantum correlations due to the high
symmetry of quantum states.

\section{Conclusions}
\label{sec:Concl}

The spin-$(j,1/2)$ model is a nontrivial example of a high-dimensional quantum system
for which nonclassical correlations, namely LQU and LQFI, can be evaluated in an exact
analytical form \cite{GTA13,GSGTFSSOA14}.
Moreover, restricting the model to U(1) axial symmetry provides compact formulas for
the discord-like correlations while preserving a large number of interactions.
This makes the obtained formulas (\ref{eq:F0})--(\ref{eq:U0U1}) an indispensable tool
when performing extensive studies of various properties of quantum correlations.

For the fully isotropic Heisenberg XXX model, we have shown that the threshold
temperature $T_{\rm th}$ of quantum entanglement (both negativity and EoF) vanishes in
the classical limit.

The following results have been obtained for the ground-state quantum correlations.
In the fully isotropic Heisenberg model [the Z$_2\times$SU(2) case], the quantum
correlations LQU and LQFI, unlike both types of entanglement ($\cal N$ and EoF), are
nonzero in the limit $j\to\infty$.
It has been proposed to interpret this phenomenon as a persistence of the quantum
nature of a system with macroscopically large spin.
We have also revealed that the quantum correlations can increase with increasing spin
$j$.

Our further analysis showed that the named macroscopic quantum correlations are
unstable and are destroyed by any appearance of external magnetic fields or violation
of the spatial isotropy of couplings.
On the other hand, in the XXZ system with dominant longitudinal interaction,
oscillations of quantum correlations LQU and LQFI arise.
We have shown that these are also instabilities and they are eliminated by an external
field.

Finally, we have found the LQFI correlation in terms of the parameter $F$ that defines
the SU(2) quantum state.
Our result [see Eq.~(\ref{eq:Fxxx-vsF}) and Fig.~\ref{fig:zXXX_LQFIvsF}] adds to the
list of achievements for this highly symmetric quantum state.

New features revealed in the behavior of quantum correlations shed additional light on
the mechanism of the transition of a quantum system into a classical counterpart with
increasing its size (spin, in the given case).

\section*{Acknowledgements}
M.A.Y. and E.I.K. were supported in part by a state task, the state
registration number of the Russian Federation is $\#$124013000760-0. S.H. thanks the School of Particles and Accelerators at the Institute for Research in Fundamental Sciences for their financial support.


\section*{Data availability statement}
All data that support the findings of this study are included within the article.


\appendix
\section{
LQU and LQFI of general spin-$(j,1/2)$ axially symmetric states
}
\label{sec:append}
The LQU and LQFI formulas for axially symmetric qubit-quDit systems have been obtained
in Ref.~\cite{HKY25}.
Here, we derive similar formulas for the quDit-qubit system, i.e. when the Hilbert
subspaces exchange places with each other.

So, let the quantum state be U(1) invariant, that is, the density matrix commutes with the
$z$-component of the total spin, given by Eq.~(\ref{eq:Sz_total}).
The corresponding density matrix is written as
\begin{equation}
   \label{eq:rho}
   \rho=
	 \left(
      \begin{array}{cccccccccccc}
      p_0&\ &\ &\ &\ &\ &\ &\ &\ &\ &\ &\ \\
      \ &a_1&u_1&\ &\ &\ &\ &\ &\ &\ &\ &\ \\
      \ &u_1^*&b_1&\ &\ &\ &\ &\ &\ &\ \ &\ &\ \\
      \ &\ &\ &\ddots&\ &\ &\ &\ &\ &\ &\ &\ \\
      \ &\ &\ &\ &a_k&u_k&\ &\ &\ &\ &\ &\ \\
      \ &\ &\ &\ &u_k^*&b_k&\ &\ &\ &\ &\ &\ \\
      \ &\ &\ &\ &\ &\ &\ &\ddots&\ &\ &\ &\ \\
      \ &\ &\ &\ &\ &\ &\ &\ &\ &a_{2j}&u_{2j}&\ \\
      \ &\ &\ &\ &\ &\ &\ &\ &\ &u_{2j}^*&b_{2j}&\ \\
			\ &\ &\ &\ &\ &\ &\ &\ &\ &\ &\ &p_{4j+1}
      \end{array}
   \right).\
\end{equation}
This matrix is Hermitian and must satisfy additional conditions $\rho\ge0$ and
${\rm Tr}\rho=1$.

The eigenvalues of the density matrix (\ref{eq:rho}) are $p_0$, $p_{4j+1}$ and plus
\begin{eqnarray}
   \label{eq:pkqk}
   &&p_k=\frac{1}{2}\Big(a_k+b_k+\sqrt{(a_k-b_k)^2+4|u_k|^2}\Big),
   \nonumber\\
   &&q_k=\frac{1}{2}\Big(a_k+b_k-\sqrt{(a_k-b_k)^2+4|u_k|^2}\Big),
\end{eqnarray}
where $k=1,\ldots,2j$.

The eigenvectors of the density matrix $\rho$ are given as
\begin{eqnarray}
   \label{eq:psi_n}
   &&|\psi_0\rangle=|e_0\rangle,
   \nonumber\\
   &&\ldots
   \nonumber\\
   &&|\psi_k\rangle={\tilde\kappa}_k|e_{2k-1}\rangle+{\tilde u}_k^*|e_{2k}\rangle,
   \nonumber\\
   &&|\varphi_k\rangle={\tilde u}_k|e_{2k-1}\rangle-{\tilde\kappa}_k|e_{2k}\rangle,
   \nonumber\\
   &&\ldots
   \nonumber\\
   &&|\psi_{4j+1}\rangle=|e_{4j+1}\rangle
\end{eqnarray}
in which
\begin{equation}
   \label{eq:e_i}
   |e_i\rangle=
	 \left(
      \begin{array}{l}
      \delta_{i,0}\\
      \delta_{i,1}\\
      \dots\\
			\delta_{i,4j+1}
      \end{array}
   \right)\qquad (i=0,1,\ldots,4j+1),
\end{equation}
where $\delta_{i,n}$ is the Kronecker delta, and
\begin{equation}
   \label{eq:kappa_tilde}
   {\tilde\kappa}_k=\kappa_k/\sqrt{\kappa_k^2+|u_k|^2},\qquad {\tilde u}_k=u_k/\sqrt{\kappa_k^2+|u_k|^2}
\end{equation}
with
\begin{equation}
   \label{eq:kappa_k}
	 \kappa_k=\frac{1}{2}\Big(a_k-b_k+\sqrt{(a_k-b_k)^2+4|u_k|^2}\Big).
\end{equation}
From here and Eq.~(\ref{eq:kappa_tilde}), it follows that
\begin{equation}
   \label{eq:kappa_ku2_k}
   {\tilde\kappa}_k^2=\frac{1}{2}\Big(1+\frac{a_k-b_k}{p_k-q_k}\Big),\qquad
	 {\tilde u}_k^2=\frac{1}{2}\Big(1-\frac{a_k-b_k}{p_k-q_k}\Big).
\end{equation}

The LQFI is given by
\begin{equation}
   \label{eq:calF}
   {\cal F}=1-\lambda_{\max}^{(M)},
\end{equation}
where $\lambda_{\max}^{(M)}$ is the largest eigenvalue of the real symmetric $3\times3$
matrix $M$ with entries
\begin{equation}
   \label{eq:M}
   M_{\mu\nu}=\sum_{m,n}\frac{2P_mP_n}{P_m+P_n}\langle\Psi_m|I_{2j+1}\otimes\sigma_\mu|\Psi_n\rangle
	 \langle\Psi_n|I_{2j+1}\otimes\sigma_\nu|\Psi_m\rangle,
\end{equation}
where $P_m+P_n\ne0$.
\\
\\
Here, $P_m$ ($m=0,1,\ldots,4j+1$) are the ordered eigenvalues of the density matrix
(\ref{eq:rho}),
\begin{equation}
   \label{eq:Pm}
   P_m=
   \begin{cases}
   p_0, &m=0,\cr
   p_k, &m=2k-1,\cr
   q_k, &m=2k,\cr
   p_{4j+1}, &m=4j+1
   \end{cases}
\end{equation}
and the corresponding eigenvectors are given as
\begin{equation}
   \label{eq:Psi_m}
   |\Psi_m\rangle=
   \begin{cases}
   |\psi_0\rangle, &m=0,\cr
   |\psi_k\rangle, &m=2k-1,\cr
   |\varphi_k\rangle, &m=2k,\cr
   |\psi_{4j+1}\rangle, &m=4j+1.
   \end{cases}
\end{equation}
Using these eigenvectors, we find that the matrix $M$ is diagonal, $M_{yy}=M_{xx}$ and
\begin{eqnarray}
   \label{eq:Mxx}
   \frac{1}{4}M_{xx}&=&\frac{p_0p_1}{p_0+p_1}{\tilde\kappa}_1^2
	 +\frac{p_0q_1}{p_0+q_1}|{\tilde u}_1|^2
	 +\frac{q_{2j}p_{4j+1}}{q_{2j}+p_{4j+1}}{\tilde\kappa}_{2j}^2\nonumber\\
	&+&\frac{p_{2j}p_{4j+1}}{p_{2j}+p_{4j+1}}|{\tilde u}_{2j}|^2
   +\sum_{k=1}^{2j-1}\Big[\frac{q_kp_{k+1}}{q_k+p_{k+1}}{\tilde\kappa}_k^2{\tilde\kappa}_{k+1}^2
   \nonumber\\
	 &+&\frac{q_kq_{k+1}}{q_k+q_{k+1}}{\tilde\kappa}_k^2|{\tilde u}_{k+1}|^2
   +\frac{p_kp_{k+1}}{p_k+p_{k+1}}|{\tilde u}_k|^2{\tilde\kappa}_{k+1}^2\nonumber\\
   &+&\frac{p_kq_{k+1}}{p_k+q_{k+1}}|{\tilde u}_k|^2|{\tilde u}_{k+1}|^2\Big]
\end{eqnarray}
and
\begin{eqnarray}
   \label{eq:Mzz}
   M_{zz}&=&p_0+p_{4j+1}
   \nonumber\\
   &+&\sum_{k=1}^{2j}\big[(p_k+q_k)({\tilde\kappa}_k^2-|{\tilde u}_k|^2)^2
	 +16\frac{p_kq_k}{p_k+q_k}{\tilde\kappa}_k^2|{\tilde u}_k|^2\big].
   \nonumber\\
\end{eqnarray}
Hence, there are two branches (sub-functions), ${\cal F}_0=1-M_{zz}$ and
${\cal F}_1=1-M_{xx}$, the minimum of which determines LQFI according to
Eq.~(\ref{eq:calF}).

Further, the LQU is given by
\begin{equation}
   \label{eq:Um}
   {\cal U}=1-\lambda_{\max}^{(W)},
\end{equation}
where $\lambda_{\max}^{(W)}$ denotes the maximum eigenvalue of the $3\times3$ symmetric
matrix $W$ whose entries are
\begin{equation}
   \label{eq:W}
   W_{\mu \nu}={\rm Tr}\{\rho^{1/2}(I_{2j+1}\otimes\sigma_\mu)\rho^{1/2}(I_{2j+1}\otimes\sigma_\nu)\}
\end{equation}
with $\mu,\nu=x,y,z$, and $\sigma_{x,y,z}$ are the Pauli matrices as before.

Using eigenvectors of the density matrix, we find that the matrix $W$ is also diagonal,
$W_{yy}=W_{xx}$ and
\begin{eqnarray}
   \label{eq:Wxx}
   \frac{1}{2}W_{xx}&=&
   {\tilde\kappa}_1^2\sqrt{p_0p_1}+|{\tilde u}_1|^2\sqrt{p_0q_1}+{\tilde\kappa}_{2j}^2\sqrt{q_{2j}p{4j+1}}
   \nonumber\\
   &+&|{\tilde u}_{2j}|^2\sqrt{p_{2j}p_{4j+1}}
   +\sum_{k=1}^{2j-1}\big({\tilde\kappa}_k^2\sqrt{q_k}
   \\
	 &+&
   |{\tilde u}_k|^2\sqrt{p_k}\big)\big({\tilde\kappa}_{k+1}^2\sqrt{p_{k+1}}+|{\tilde u}_{k+1}|^2\sqrt{q_{k+1}}\big)
   \nonumber
\end{eqnarray}
and
\begin{eqnarray}
   \label{eq:Wzz}
   W_{zz}&=&p_0+p_{4j+1}
   \\
	 &+&\sum_{k=1}^{2j}\big[(p_k+q_k)({\tilde\kappa}_k^2-|{\tilde u}_k|^2)^2
	 +8{\tilde\kappa}_k^2|{\tilde u}_k|^2\sqrt{p_kq_k}\big].
   \nonumber
\end{eqnarray}
So, again, there are two branches ${\cal U}_0=1-W_{zz}$ and ${\cal U}_1=1-W_{xx}$,
the minimum of which determines LQU in accord with Eq.~(\ref{eq:Um}).

Substituting Eq.~(\ref{eq:kappa_ku2_k}) into Eqs.~(\ref{eq:Mzz}), (\ref{eq:Wxx}) and
(\ref{eq:Wzz}), we come to surprisingly compact formulas for the branches
\begin{equation}
   \label{eq:F0}
   {\cal F}_0=4\sum_{k=1}^{2j}\frac{|u_k|^2}{a_k+b_k},
\end{equation}
\begin{equation}
   \label{eq:U0}
   {\cal U}_0=4\sum_{k=1}^{2j}\frac{|u_k|^2}{a_k+b_k+2\sqrt{a_kb_k-|u_k|^2}}
\end{equation}
and
\begin{eqnarray}
   \label{eq:U1}
   {\cal U}_1&=&1-2\Bigg[\frac{a_1+\sqrt{p_1q_1}}{\sqrt{p_1}+\sqrt{q_1}}\sqrt{p_0}
	 +\frac{b_{2j}+\sqrt{p_{2j}q_{2j}}}{\sqrt{p_{2j}}+\sqrt{q_{2j}}}\sqrt{p_{4j+1}}
\nonumber\\
   &+&\sum_{k=1}^{2j-1}\frac{(b_k+\sqrt{p_k q_k})(a_{k+1}+\sqrt{p_{k+1}q_{k+1}})}
	 {(\sqrt{p_k}+\sqrt{q_k})(\sqrt{p_{k+1}}+\sqrt{q_{k+1}})}
	 \Bigg].
\end{eqnarray}

Similarly, the  final formula for the branch ${\cal F}_1$ is written as
\begin{eqnarray}
   \label{eq:F1a}
   {\cal F}_1&=&1-4\Bigg(\frac{p_0(a_1p_0+p_1q_1)}{(p_0+p_1)(p_0+q_1)}
	 +\frac{p_{4j+1}(b_{2j}p_{4j+1}+p_{2j}q_{2j})}{(p_{2j}+p_{4j+1})(q_{2j}+p_{4j+1})}\Bigg)
   \nonumber\\
	 &-&\sum_{k=1}^{2j-1}\Bigg[
	 \frac{q_k(a_{k+1}q_k+p_{k+1}q_{k+1})}{(q_k+p_{k+1})(q_k+q_{k+1})}\bigg(1+\frac{a_k-b_k}{p_k-q_k}\bigg)\nonumber\\
     &+&\frac{p_k(a_{k+1}p_k+p_{k+1}q_{k+1})}{(p_k+p_{k+1})(p_k+q_{k+1})}\bigg(1-\frac{a_k-b_k}{p_k-q_k}\bigg)
   \nonumber\\
   &+&
	\frac{p_{k+1}(b_kp_{k+1}+p_kq_k)}{(p_k+p_{k+1})(q_k+p_{k+1})}\bigg(1+\frac{a_{k+1}-b_{k+1}}{p_{k+1}-q_{k+1}}\bigg)\nonumber\\
	 &+&\frac{q_{k+1}(b_kq_{k+1}+p_kq_k)}{(p_k+q_{k+1})(q_k+q_{k+1})}\bigg(1-\frac{a_{k+1}-b_{k+1}}{p_{k+1}-q_{k+1}}\bigg)
	 \Bigg].
\end{eqnarray}
%
After some algebra, we get
\begin{widetext}
\begin{eqnarray}
   \label{eq:F1}
   {\cal F}_1&=&1-4\Bigg(\frac{p_0(a_1p_0+p_1q_1)}{(p_0+p_1)(p_0+q_1)}
	 +\frac{p_{4j+1}(b_{2j}p_{4j+1}+p_{2j}q_{2j})}{(p_{2j}+p_{4j+1})(q_{2j}+p_{4j+1})}\Bigg)
	 -4\sum_{k=1}^{2j-1}\bigg[a_{k+1}b_k(a_k+a_{k+1})(b_k+b_{k+1})(a_k+b_{k+1})
   \nonumber\\
   &-&a_{k+1}|u_k|^2\big(b_{k+1}^2+b_k(2a_k+a_{k+1})+b_{k+1}(a_k+b_k)\big)
   -b_k|u_{k+1}|^2\big(a_k^2+a_{k+1}(b_k+2b_{k+1})+a_k(a_{k+1}+b_{k+1})\big)
   \nonumber\\
   &+&(a_k+b_{k+1})|u_k|^2|u_{k+1}|^2+a_{k+1}|u_k|^4+b_k|u_{k+1}|^4
	 \bigg]/\big[(p_k+p_{k+1})(p_k+q_{k+1})(q_k+p_{k+1})(q_k+q_{k+1})\big].
\end{eqnarray}
\end{widetext}
Both of these formulas are suitable for numerical calculations of ${\cal F}_1$.

This makes it possible to actually calculate the nonclassical correlations LQFI
\begin{equation}
   \label{eq:F0F1}
   {\cal F}=\min{\{{\cal F}_0,{\cal F}_1\}}
\end{equation}
and LQU
\begin{equation}
   \label{eq:U0U1}
   {\cal U}=\min{\{{\cal U}_0,{\cal U}_1\}}.
\end{equation}

Thereby, the derived formulas for the branches of discord-type quantum correlations are
simple and very convenient for performing calculations on a computer.









\end{document}